\documentclass[12pt]{iopart}
\usepackage{epsfig}

\begin{document}

\title[Jets in deep-inelastic scattering at HERA]{Jets in deep-inelastic scattering at HERA}

\author{Matthew Wing\footnote{\tt wing@mail.desy.de} \\
(on behalf of the H1 and ZEUS collaborations)
}

\address{University of Bristol, DESY, Notkestrasse 85, 22607 Hamburg, Germany}

\begin{abstract}
Jet cross sections in deep-inelastic scattering over a wide region of phase space have 
been measured at HERA. These cross section measurements provide a thorough test of the 
implementation of Quantum Chromodynamics in next-to-leading order (NLO) calculations. They also 
provide the opportunity to test the consistency of the gluon distribution in the proton 
as extracted from (mainly) inclusive DIS measurements. Comparison of the cross sections with 
NLO enables accurate extractions of the strong coupling constant, $\alpha_s$, to be made, 
several of which are reported here.
\end{abstract}




\section{Introduction}
\label{sec:intro}

Deep-inelastic $ep$ scattering (DIS) at HERA, in which a photon of virtuality, $Q^2$, is 
exchanged, can lead to the production of jets in the final state. Cross-section measurements 
of these types of processes provide stringent tests of the perturbative Quantum 
Chromodynamics (pQCD) formalism in next-to-leading order (NLO) calculations, the structure of the 
proton and allow measurements of the strong coupling constant, $\alpha_s$. These features can 
be seen in the formula for the cross section, $d\sigma$, which is factorised into a 
convolution of the hard partonic cross section, 
$d\hat{\sigma}_a (x,\alpha_s(\mu_R^2),\mu_R^2, \mu_F^2)$, and the proton's parton density, 
$f_a(x,\mu_F^2)$:
\begin{equation}
d\sigma = \sum_{a=q,\bar{q},g}\int dx \ f_a(x,\mu_F^2) \ 
          d\hat{\sigma}_a (x,\alpha_s(\mu_R),\mu_R^2, \mu_F^2).
\end{equation}
The hard partonic cross section is a power series expansion in $\alpha_s$ and is calculable 
in pQCD. The proton's parton density is derived from fits to published data, such as 
inclusive DIS measurements~\cite{f2_pdf} and high transverse energy jet production in 
$p\bar{p}$ production~\cite{jet_pdf}.

\begin{figure}[htp]
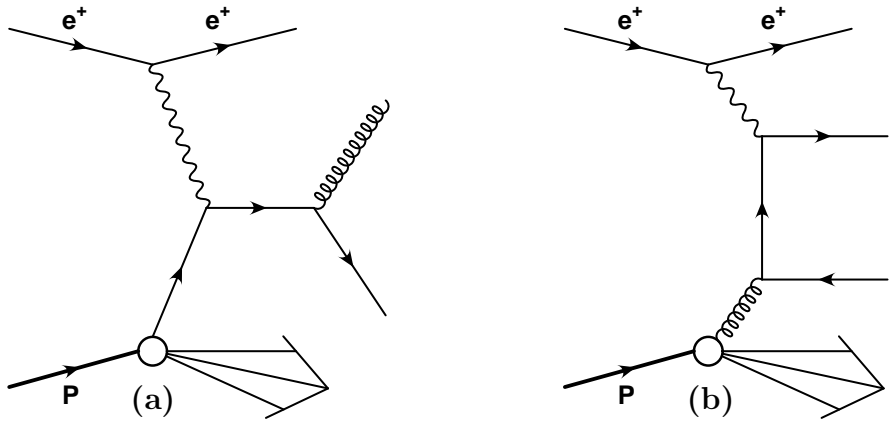

\caption{\label{fig:feyn}(a) The QCD Compton and (b) boson-gluon fusion processes.}
\end{figure}

In the region of momentum fraction, $x$, and $Q^2$ in which DIS jet measurements are 
currently performed, the quark density in the photon is well constrained from particularly 
inclusive DIS measurements. At high $Q^2$ where the quark initiated process, QCD Compton 
(see figure~\ref{fig:feyn}(a)), is dominant, the jet cross section measurements test 
pQCD. At lower $Q^2$, where the boson-gluon fusion process (see figure~\ref{fig:feyn}(b)) is 
dominant, measurements of the jet cross sections provide complementary information on the 
gluon density in the proton, which from DIS is only constrained through the interpretation 
of scaling violations in NLO fits.

\begin{center}
~\epsfig{file=wing.hpp-comp.epsi,height=5.5cm}
\hspace{2cm}~\epsfig{file=wing.hpp-bg.epsi,height=5.5cm}
\end{center}
\unitlength=1mm
\begin{picture}(0,0)(100,100)
\put(137,110){\bf (a)}
\put(211,110){\bf (b)}
\end{picture}

In this paper, the latest measurements from both the H1 and ZEUS collaborations on DIS 
jet production will be discussed. Initially the jet cross section comparisons with NLO 
will be shown and then the extraction of the gluon density and $\alpha_s$ will be presented. 

Limitations of both the current data and theory and some of their possible solutions will 
be addressed. In particular, the measurements raise the following issues:

\begin{itemize}

\item at which points are the theoretical or experimental errors dominant?

\item which of the two ``natural'' scales, $Q$ and transverse energy, $E_T$, is the more 
      appropriate?

\item is there a region where the DGLAP~\cite{dglap} formalism breaks down and 
      BFKL~\cite{bfkl} is more appropriate?

\item is there an effect of the resolved photon?

\end{itemize}

\section{Inclusive jet measurements}

The two inclusive jet measurements presented here have different goals; the first is 
concentrated in a ``safe'' region to test pQCD and extract $\alpha_s$, whilst the second 
extends the kinematic phase space to more extreme regions to look for suggestions of BFKL 
effects. The safe region was defined by considering jets at a reasonably 
high$-Q^2$ of greater than 125~GeV$^2$, high$-E_{T, \rm jet}^{\rm B}$ of greater than 8 GeV 
in a central region of the detector, $-2<\eta_{\rm jet}^{\rm B}<1.8$, where the jets are 
reconstructed in the Breit frame of reference. To be more sensitive 
to BFKL effects, a second region was probed; $5<Q^2<100$~GeV$^2$, 
$E_{T, \rm jet}^{\rm B}>8$ GeV and $-1<\eta^{\rm LAB}_{\rm jet}<2.8$.

The measured cross section as a function of $Q^2$, for $Q^2>125$~GeV$^2$, is shown in 
figure~\ref{fig:incl}(a) compared to predictions from NLO QCD implemented in the 
{\sc Disent}~\cite{disent}
program. The data fall by five orders of magnitude and are reasonably well described by 
the predictions corrected for hadronisation effects and using two different values for the 
renormalisation scale, $\mu_R$. The description is quantified in figure~\ref{fig:incl}(b) 
where the ratio of the data to the prediction (using $\mu_R=Q$) is shown. The data lie 
above the theory by about $12\%$ at $Q^2<500$~GeV$^2$, although the size of the theoretical 
and experimental uncertainties rule out any firm conclusions. At higher $Q^2$, the prediction 
describes the data well. Similar results were observed when considering the cross section 
as a function of the transverse energy of the jet. The data can therefore be used to extract 
a value of $\alpha_s$, which was done at high$-Q^2$ and 
high$-E_{T, \rm jet}^{\rm B}$. As can be seen in figure~\ref{fig:incl}(b), extracting $\alpha_s$ 
at $Q^2>500$~GeV$^2$ leads to a smaller theoretical uncertainty, arising mainly from the 
variation of the renormalisation scale by a factor of two.

\begin{center}
~\epsfig{file=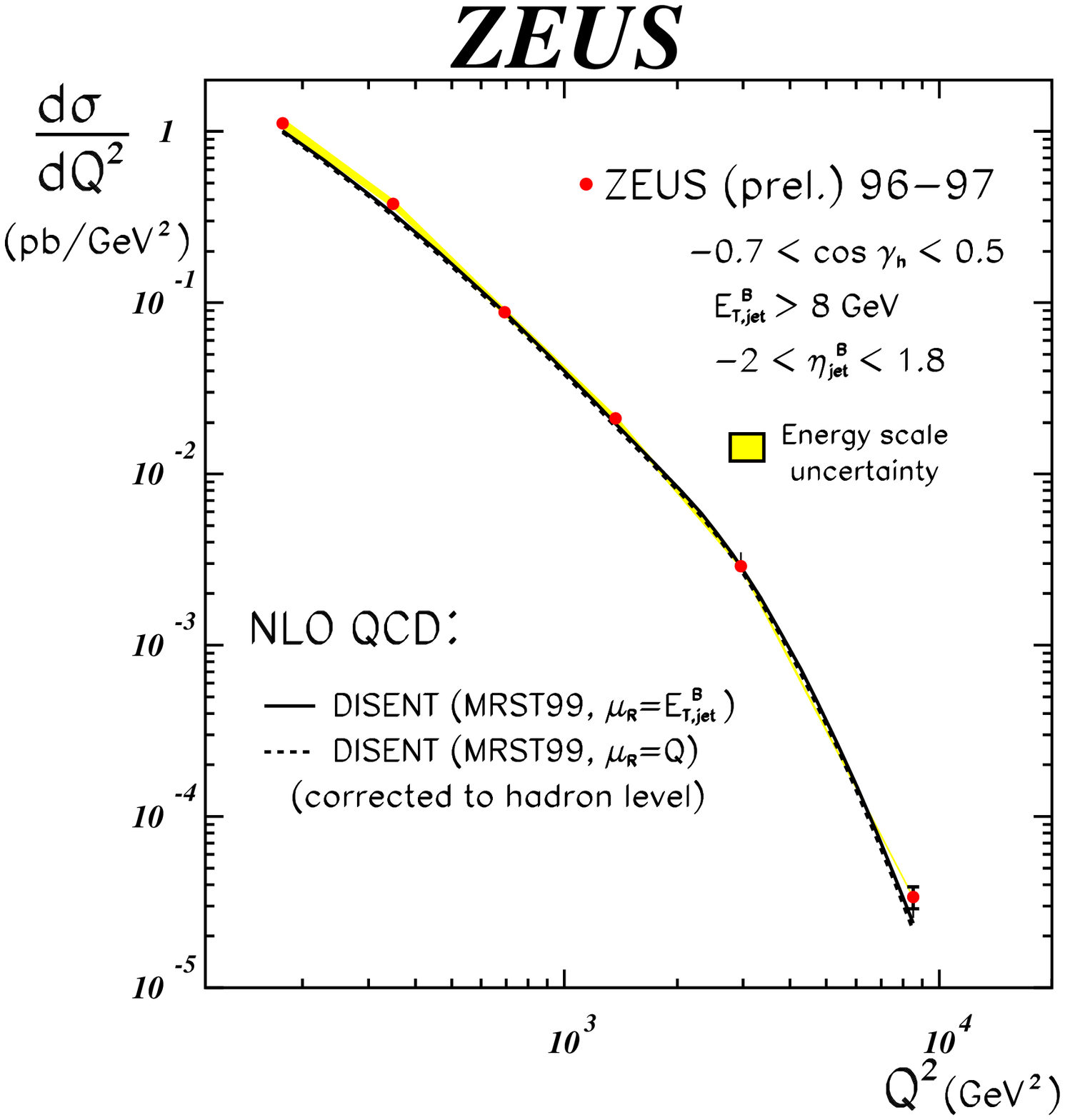,width=7.65cm}
~\epsfig{file=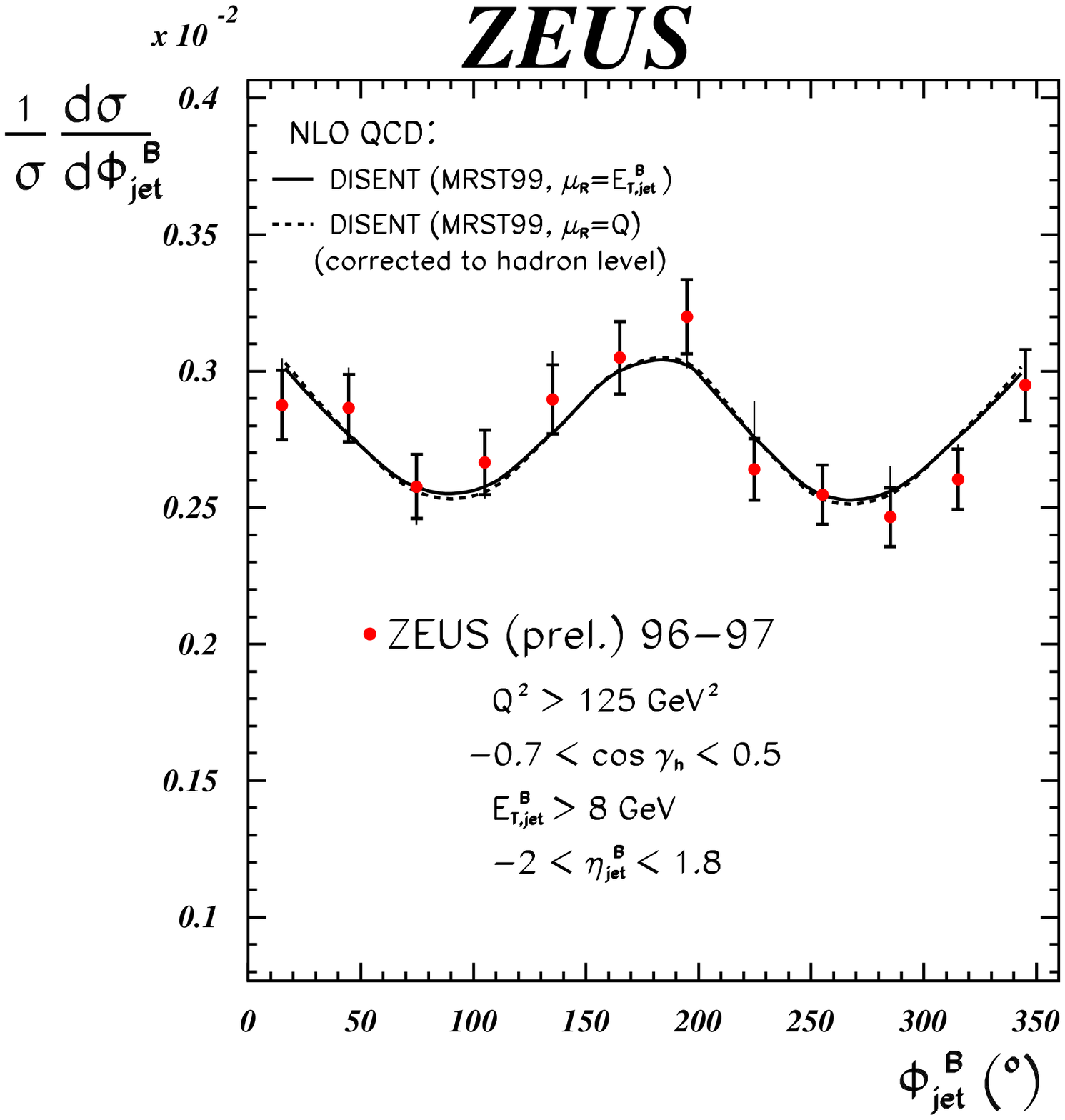,height=7.65cm}
\end{center}
~\epsfig{file=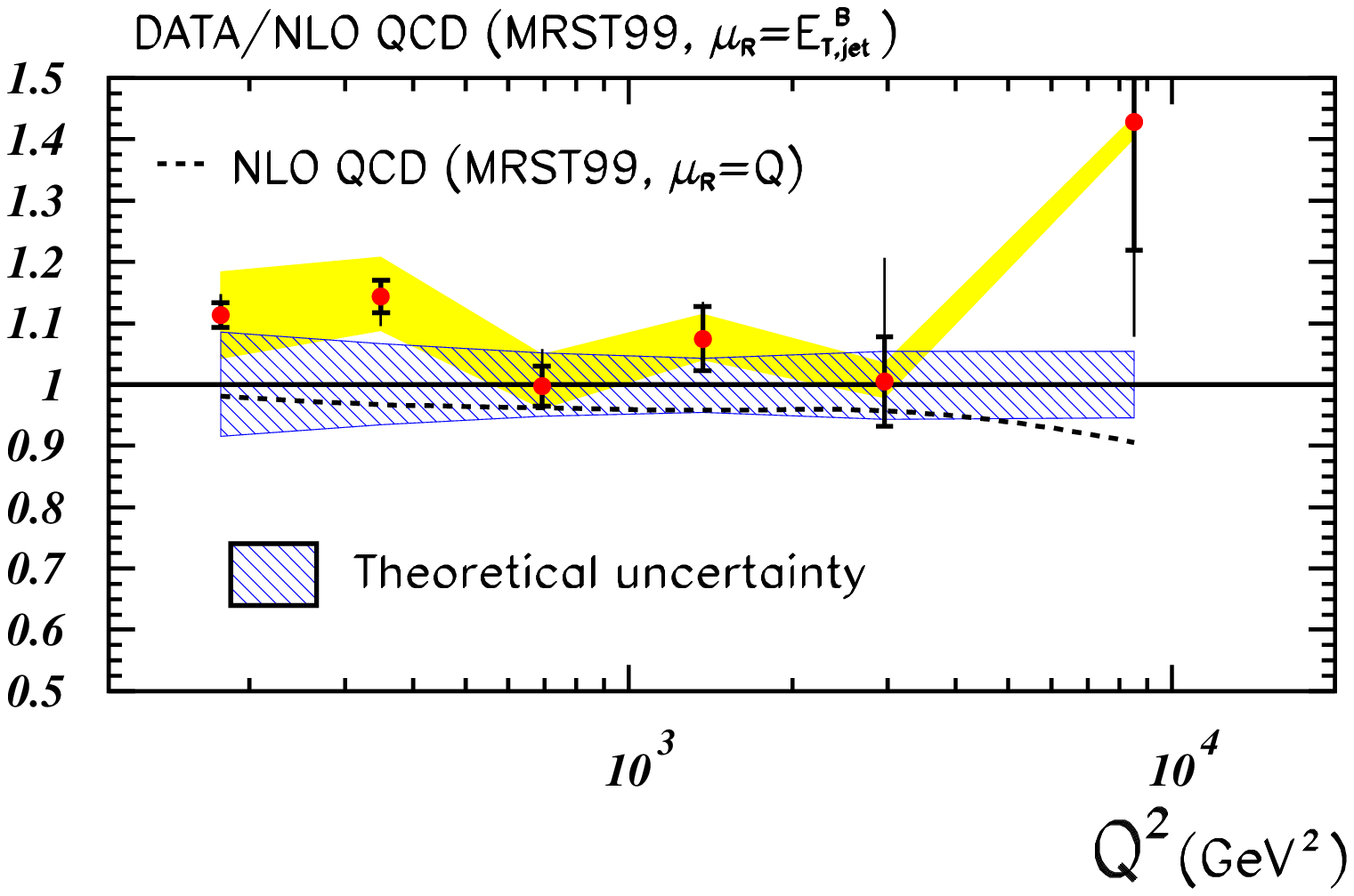,width=7.65cm}
\unitlength=1mm
\begin{picture}(0,0)(100,100)
\put(83,160){\bf (a)}
\put(83,115){\bf (b)}
\put(162,160){\bf (c)}
\end{picture}
\begin{figure}[htp]
\caption{\label{fig:incl}Inclusive jet cross sections as a function of (a) $Q^2$ and (b) 
the ratio of data to NLO and (c) the normalised cross section as a function of 
$\phi^{\rm B}_{\rm jet}$. The data are shown as the points with statistical errors (inner 
bars) and statistical and systematic errors added in quadrature (outer bars). The NLO 
is shown with two different renormalisation scales; $E_{T, \rm jet}^{\rm B}$ (solid line) 
and $Q$ (dashed line).}
\end{figure}

A further test of QCD is the structure of the cross section in the azimuthal angle, 
$\phi_{\rm jet}^{\rm B}$, which is defined as the angle between the lepton scattering plane 
and the jet production plane. The cross section is predicted to have the form~\cite{phi_form},
\begin{equation}
\frac{d\sigma}{d\phi_{\rm jet}^{\rm B}} = A + B \cos  \phi_{\rm jet}^{\rm B}
                                            + C \cos 2\phi_{\rm jet}^{\rm B},
\end{equation}
and has not be seen before in neutral-current DIS jet production. The effect has been seen 
in charged hadron production in DIS~\cite{had_phi}. Without tagging the nature of the final 
state jets, the cross section is expected to reduce to:
\begin{equation}
\frac{d\sigma}{d\phi_{\rm jet}^{\rm B}} = A +  C \cos 2\phi_{\rm jet}^{\rm B}.
\label{eq:phi_red}
\end{equation}
The measurement is shown in figure~\ref{fig:incl}(c) compared to the NLO predictions. The 
form of Eq.~(\ref{eq:phi_red}) is clearly seen in the data for the first time in neutral 
current DIS jet production and the calculation gives a good description of the shape.

Comparisons of inclusive jet data at lower scales and for more forward-going jets with 
NLO are shown in figure~\ref{fig:incl_forward}. The data are shown in different regions 
of pseudorapidity of the jet so as to maximise sensitivity to BFKL effects which are 
expected to be largest at forward values. The NLO calculation has been evaluated using 
two different choices of scale; $E_T$ and $Q$, shown in figures~\ref{fig:incl_forward}(a) 
and~\ref{fig:incl_forward}(b), respectively.

\begin{center}
~\epsfig{file=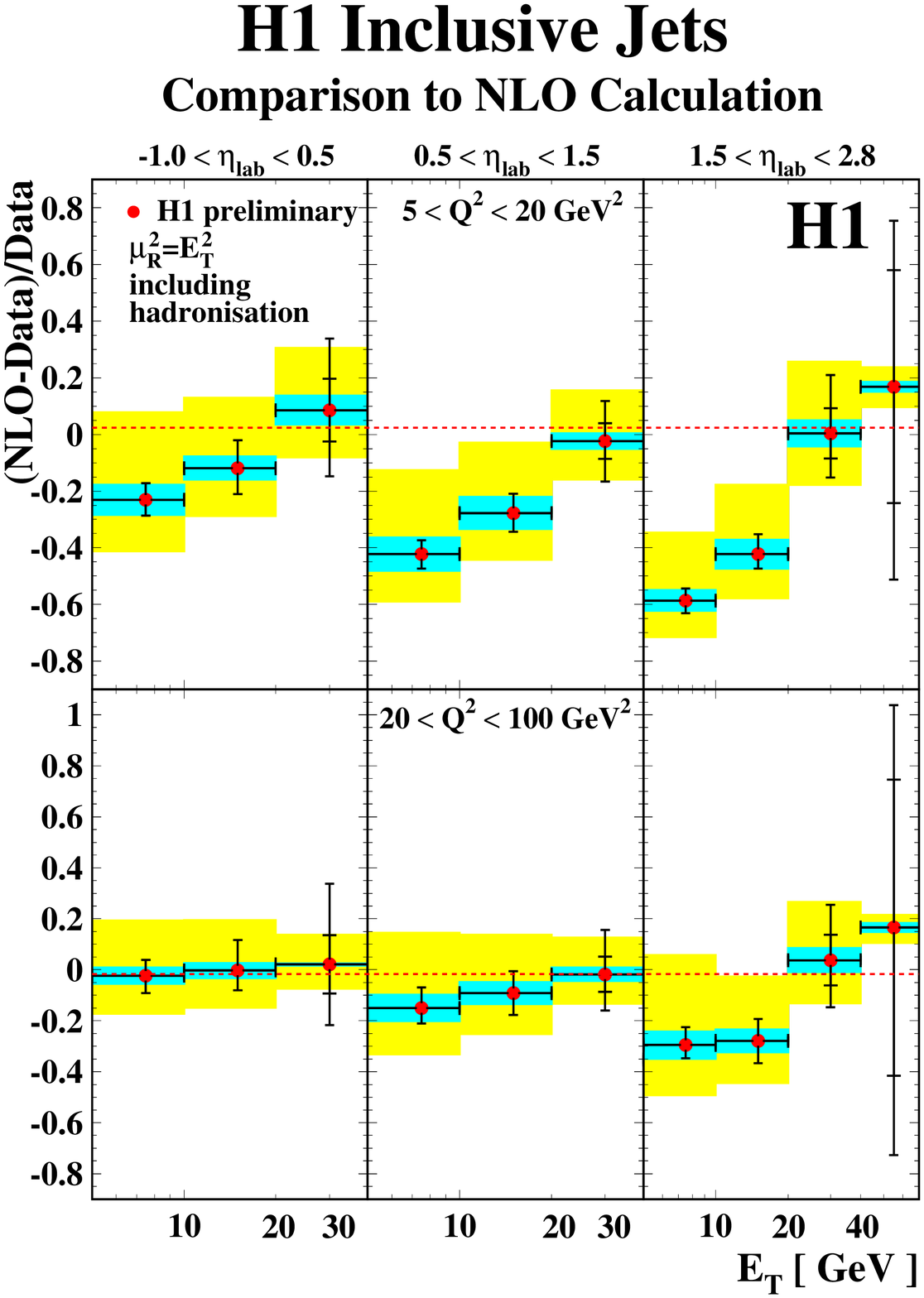,height=9cm}
~\epsfig{file=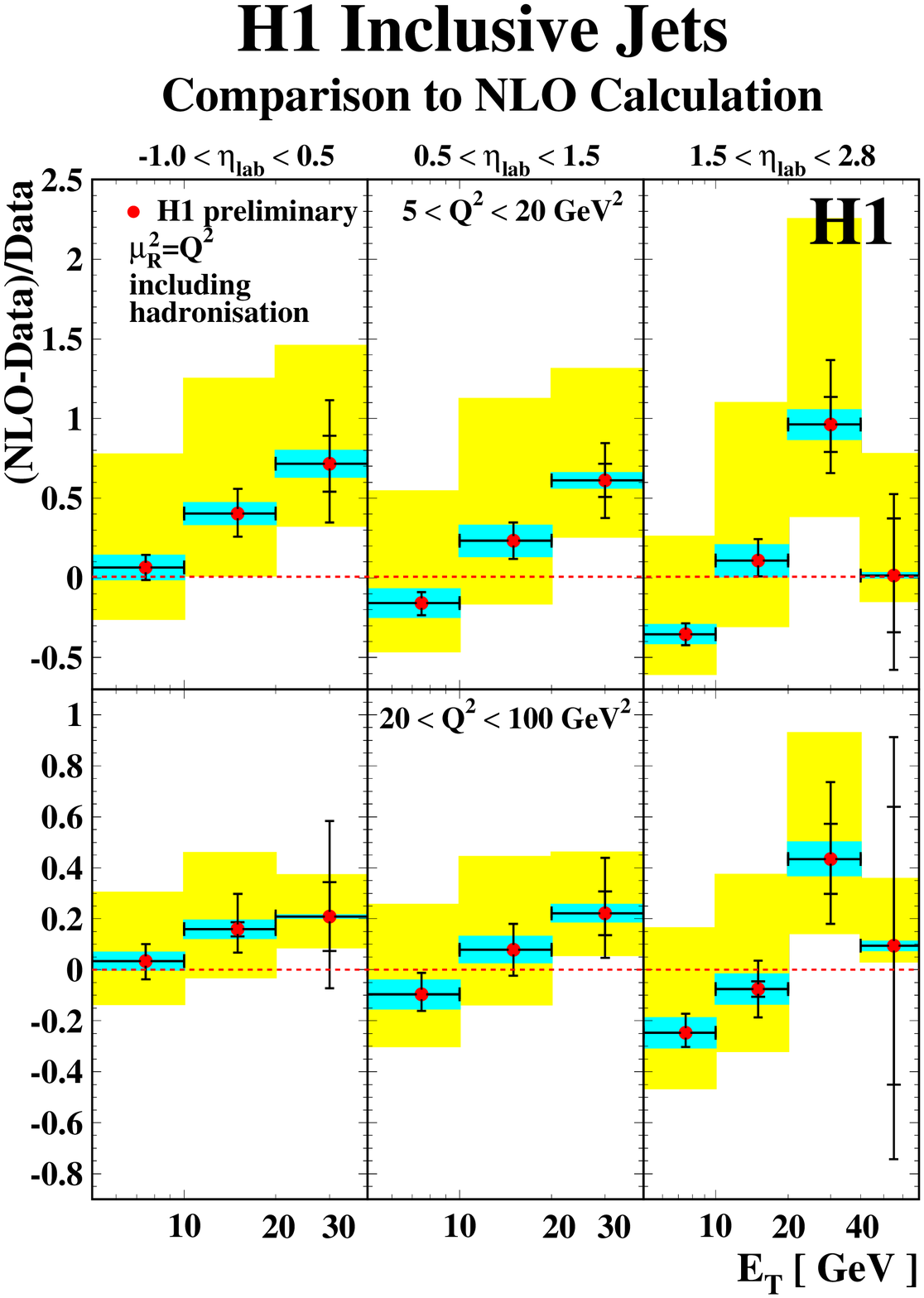,height=9cm}
\end{center}
\begin{picture}(0,0)(100,100)
\put(166,122){\bf (a)}
\put(232,122){\bf (b)}
\end{picture}
\vspace{-1cm}
\begin{figure}[htp]
\caption{\label{fig:incl_forward} Difference between data and NLO as a function of the 
transverse jet energy, when using the renormalisation scale set to (a) $E_T$ and (b) $Q$. 
The data are shown as the points with statistical errors (inner 
bars) and statistical and systematic uncertainties added in quadrature (outer bars). The NLO is 
displayed with the hadronisation (inner band) and renormalisation scale (outer band) 
uncertainties.}
\end{figure}\\
Overall, the predictions generally describe the data, although with very large uncertainties of 
sometimes up to $50-100\%$. The agreement is better at higher $Q^2$ and at lower $Q^2$ it is 
better when using $Q$ rather than $E_T$ as the scale in the calculation. It should be noted that 
at these low values of $Q^2$, $E_T >> Q$. When using $E_T$ as the scale, the NLO does 
poorly in the forward region at low values of transverse energy. This is exactly the region 
where BFKL effects are expected to show up. However, the large uncertainties on the DGLAP-based 
calculation shown, preclude any statement of the need for BFKL effects in the measurement. These 
large uncertainties in the theory, and hence a significant difference between NLO and the next 
higher order, are probably not surprising given the LO and NLO cross sections differ by up to a 
factor of ten at low $E_T$ and forward $\eta_{\rm jet}^{\rm LAB}$. Clearly this kind of 
measurement needs more accurate calculations in order to search for a breakdown of the 
DGLAP formalism.

\section{Dijet measurements}

\begin{figure}[hbp]
\begin{center}
~\epsfig{file=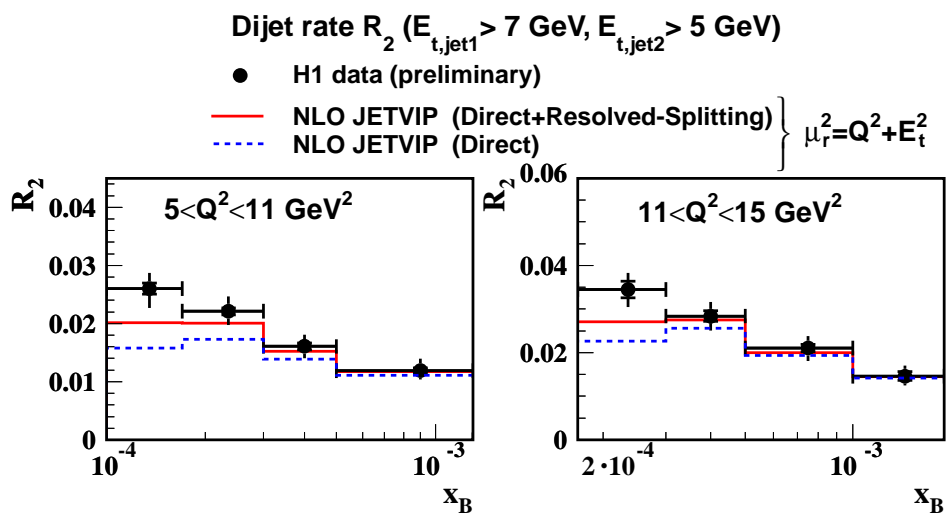,height=6.8cm}
\end{center}
\vspace{-0.5cm}
\caption{\label{fig:dijets_jetvip} Ratio of cross sections for dijet and inclusive production 
as a function of $x_{\rm B}$. The data are shown as in figure~\ref{fig:incl_forward}. The NLO 
calculation is shown with (solid line) and without (dashed line) a resolved component.}
\end{figure}

Along with BFKL effects, at low$-Q^2$ the existence and nature of a resolved photon may also 
play a r\^{o}le. The resolved photon is expected to be important when the transverse energy 
of the outgoing jets, $E_T$ is much greater than $Q$, whereas BFKL effects are expected to be 
most significant when the two quantities are roughly equal. The r\^{o}le of the resolved photon 
has been studied in dijet production with $5<Q^2<15$~GeV$^2$ and at least 
two jets, such that again $E_T >> Q$. The ratio of the dijet data as a function of the momentum 
fraction, $x_{\rm B}$, is poorly described by the calculation when the renormalisation scale is 
set to the intuitive hard scale, $\sim E_T$. When choosing the scale to be $Q$, the prediction is 
higher and describes the data well, but suffers from a lack of  predictive power due to its large 
uncertainty. Therefore, using the natural scale, the impact of the resolved photon was studied 
as implemented in the calculation {\sc JetVip}~\cite{jetvip}. The inclusion of a resolved photon 
into the calculation compared to the data is shown in figure~\ref{fig:dijets_jetvip}.
It can be seen that the calculation lies below the data 
at low$-x_{\rm B}$ and that the description improves with the inclusion of a resolved photon. 
The effect is larger at the lower range in $Q^2$, but is not enough to completely describe the 
data. Due to large uncertainties in the calculations and unknowns in the underlying physics 
processes, fundamental parameters such as $\alpha_s$, are measured at higher values of $Q^2$ 
or $E_T$, where the uncertainties are reduced.

\begin{figure}[hbp]
\begin{center}
~\epsfig{file=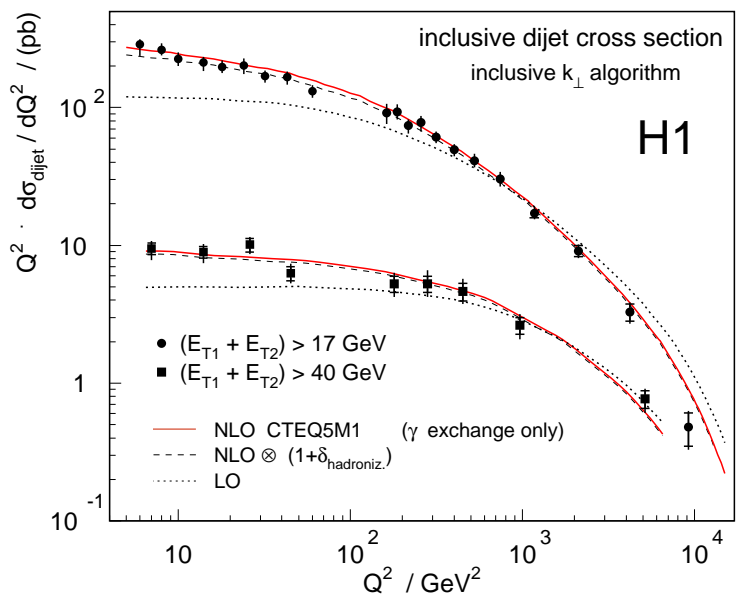,height=8cm}
\end{center}
\caption{\label{fig:dijets_h1_q2} Inclusive dijet cross section as a function of $Q^2$ for two 
regions of the sum of the transverse energies of the highest $E_T$ jets.}
\end{figure}

The dijet cross section at slightly higher $E_T$ is shown as a function of $Q^2$ in 
figure~\ref{fig:dijets_h1_q2}. 
Note that here the jets are reconstructed in the Breit frame 
whereas in figure~\ref{fig:dijets_jetvip} they were in the hadronic centre-of-mass frame. The 
description of the data by the NLO calculation in figure~\ref{fig:dijets_h1_q2} is good over 
the whole range in $Q^2$. Both data and theory also scale similarly with different minimum 
transverse energy requirements. The size of the LO to NLO corrections are reasonably large at 
low$-Q^2$ and the hadronisation corrections are also significant. These factors again demonstrate 
the need to consider higher values of $Q^2$ for the extraction of fundamental parameters. The 
description of other variables (not shown) by the NLO is also generally good.

\begin{figure}[hbp]
\begin{center}
~\epsfig{file=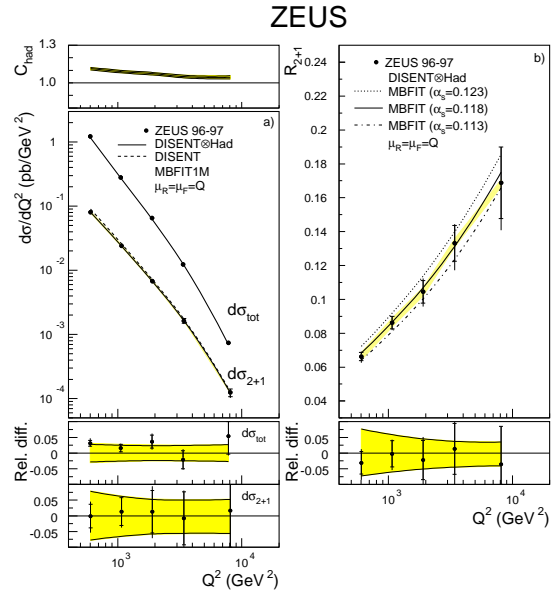,height=8cm}
\end{center}
\caption{\label{fig:dijets_zeus_q2} (a) Inclusive and dijet cross sections in DIS as a function 
of $Q^2$ and (b) the ratio of the two. The data are compared to NLO and the relative difference 
between them is also shown.}
\end{figure}

\begin{figure}[hbp]
\begin{center}
~\epsfig{file=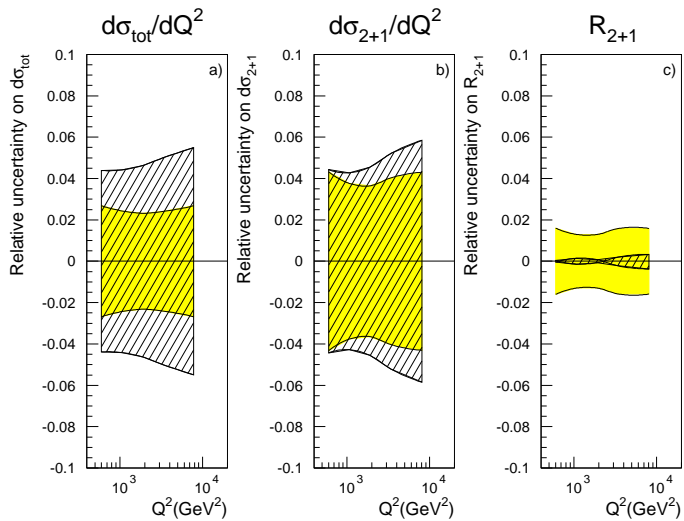,height=7.5cm}
\end{center}
\caption{\label{fig:dijets_pdf} The relative uncertainty on a) the inclusive, b) the dijet 
differential cross sections and c) the dijet fraction, due to the statistical and systematic 
uncertainties of each data set used in the determination of the MBFIT PDFs. The shaded 
and hatched bands indicate the uncertainties obtained taking into account and not taking into 
account the correlations among the PDF's parameters.}
\end{figure}

The dijet cross section at high-$Q^2$, greater than 470~GeV$^2$, is shown in 
figure~\ref{fig:dijets_zeus_q2} along with the inclusive cross section and dijet rate. All 
measurements are well described by the NLO calculation. The dijet rate is a particularly good 
variable to measure as it is sensitive to $\alpha_s$ (as demonstrated in 
figure~\ref{fig:dijets_zeus_q2}(b)) and experimental and theoretical uncertainties are expected 
to somewhat cancel. Cancellation of some of the theoretical uncertainty can be seen in 
figure~\ref{fig:dijets_pdf} where the size of the uncertainty related to the parton density 
function is shown. This was estimated using the program {\sc Epdflib}~\cite{botje}. This 
program propagates the statistical and systematic uncertainties of each data set used in the PDF, 
MBFIT and provides additional PDF sets to quantify the theoretical uncertainties. 
Figure~\ref{fig:dijets_pdf} shows that the total uncertainty arising 
from the PDF for the rate is $1.5\%$, smaller than for both the inclusive ($2.5\%$) and 
dijet ($4\%$) cross sections.

\section{Measurements of three jet production}

Neutral current DIS with three jets in the final state has recently been measured which, with 
a more complex final state, has the potential to provide a more stringent test of pQCD. A new 
NLO three-jet calculation~\cite{nlo_3jet} allows the comparison to be made for the first time. 
The ratio of three- to two-jet cross sections can also be measured. As with the ratio of cross 
sections of two-jet and inclusive production, it is expected that there will be cancellation of 
some experimental and theoretical uncertainties. Again, this measurement should allow an 
accurate extraction of $\alpha_s$, with much reduced uncertainties.

In figures~\ref{fig:3jets_xsec} and~\ref{fig:3jets_ratio}(a), the cross-section measurements 
are shown as a function of Bjorken-x, $x_{\rm Bj}$, three-jet mass, $M_{\rm 3jet}$ and $Q^2$. 
Predictions to LO and NLO, both corrected for hadronisation effects, are compared to the 
measurements. The addition of the NLO corrections are significant, particularly at 
low-$x_{\rm Bj}$ and $Q^2$ and low mass, and their inclusion into the prediction provides a 
good description of the data. Also 
shown, in figure~\ref{fig:3jets_ratio}(b), is the ratio of data to theory as a function of 
$Q^2$. The NLO can be seen to describe the data to within about $10\%$ over the whole region 
in $Q^2$. The uncertainty due to varying the value of the renormalisation scale is up 
to $30\%$ at low-$Q^2$ and decreases with increasing $Q^2$. The uncertainty arising from 
the value of $\alpha_s$ and the gluon content of the proton are reasonably constant with 
$Q^2$ and are about $20\%$ and $10\%$, respectively.

\begin{center}
~\epsfig{file=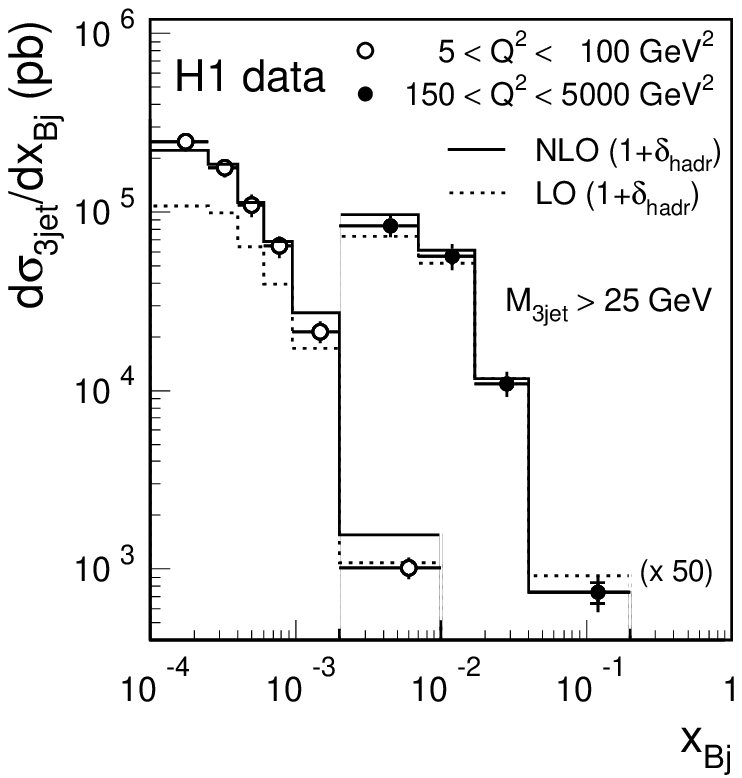,height=7.5cm}
~\epsfig{file=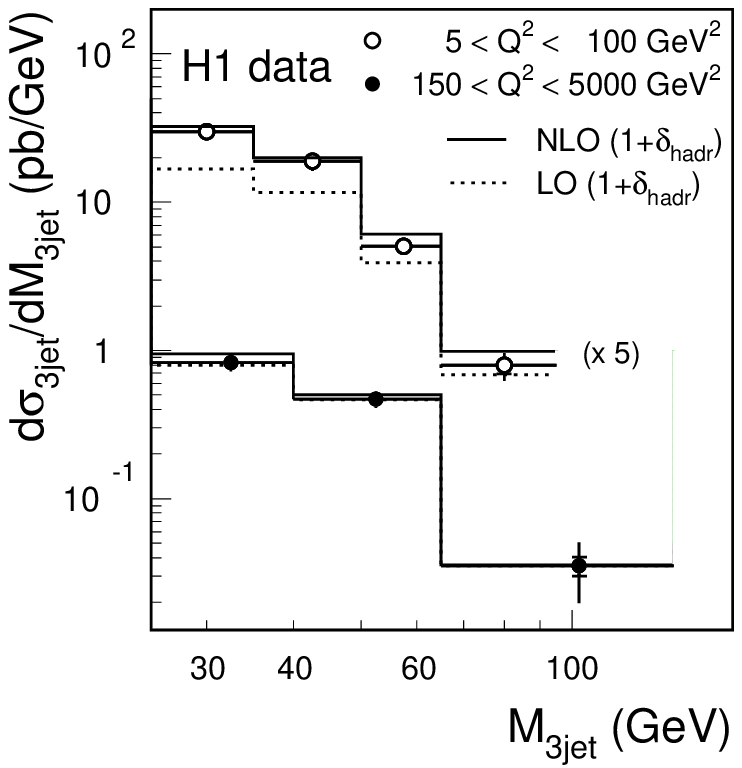,height=7.5cm}
\end{center}
\unitlength=1mm
\begin{picture}(0,0)(100,100)
\put(167,140){\bf (a)}
\put(242,140){\bf (b)}
\end{picture}
\begin{figure}[hbp]
\vspace{-1.cm}
\caption{\label{fig:3jets_xsec} The inclusive three-jet cross section measured as a function 
of (a) $x_{\rm Bj}$ and (b) $M_{\rm 3jet}$. The predictions of QCD are shown at LO (dotted line) 
and NLO (solid line), both corrected for hadronisation effects.}
\end{figure}

The ratio of three- to two-jet cross sections is also shown in figure~\ref{fig:3jets_ratio}(c)  
as a function of $Q^2$. The NLO again describes the data well and, as expected, the 
theoretical uncertainties are significantly reduced. In particular, the uncertainty due 
to the renormalisation scale and gluon PDF are both considerably smaller than the 
uncertainty arising from varying $\alpha_s$. This demonstrates the potential of the 
distribution in providing an accurate measurement of $\alpha_s$. This measurement awaits 
an increased event sample to realise its full potential.

\begin{center}
~\epsfig{file=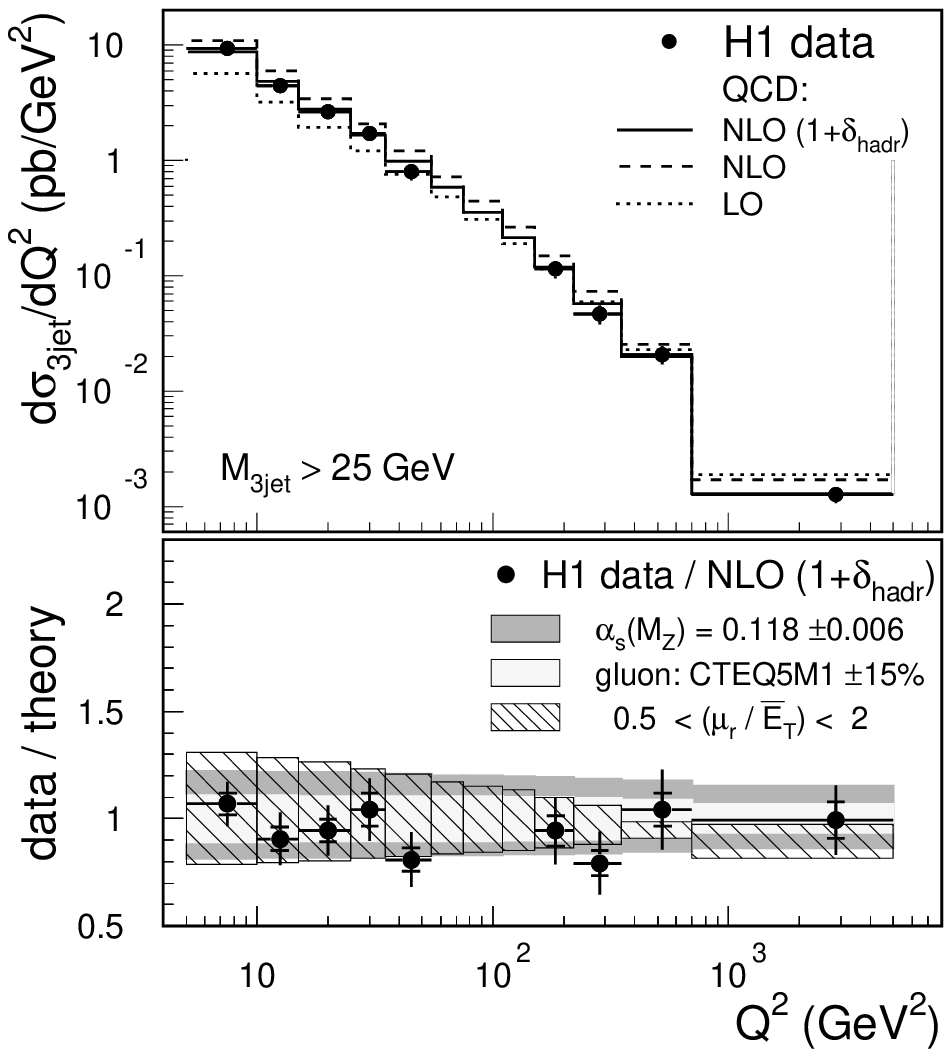,width=8cm}
~\epsfig{file=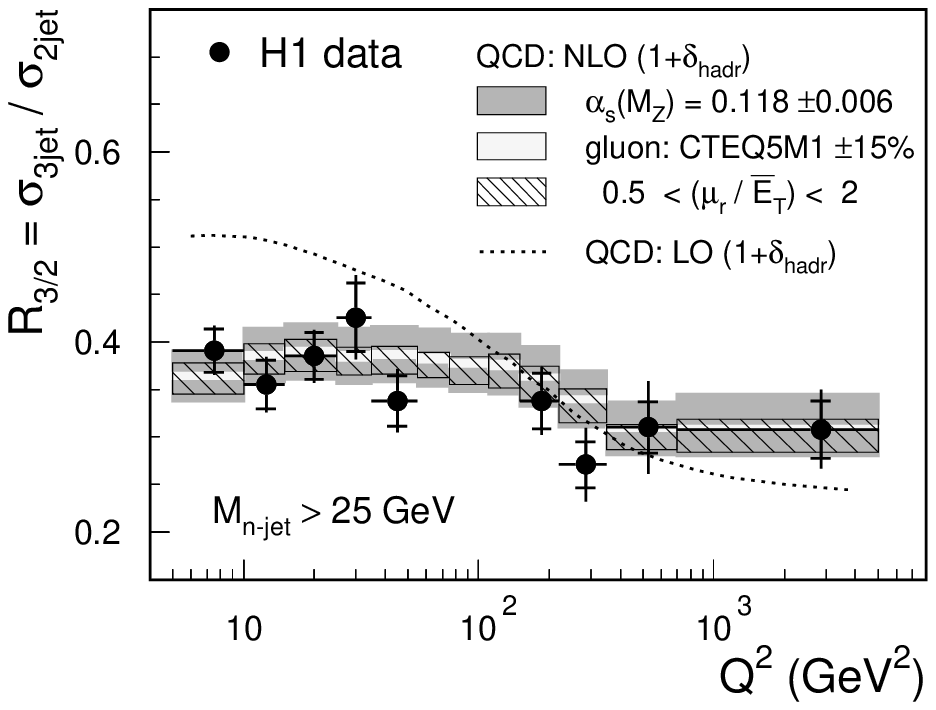,width=8cm}
\end{center}
\unitlength=1mm
\begin{picture}(0,0)(100,100)
\put(157,235){\bf (a)}
\put(157,200){\bf (b)}
\put(157,150){\bf (c)}
\end{picture}
\begin{figure}[hbp]
\vspace{-1.cm}
\caption{\label{fig:3jets_ratio} (a) The inclusive three-jet cross section measured as a 
function of $Q^2$. The predictions of QCD are shown at LO (dotted line) and NLO (solid line), 
both corrected for hadronisation effects. (b) The ratio of the cross sections in data and 
theory and the associated theoretical uncertainties. (c) The ratio of the three- to two-jet 
cross section and the associated theoretical uncertainties.}
\end{figure}

\section{Measurements of jet substructure}

Measurements of jet substructure in NC DIS have been performed and used to test pQCD and 
perform extractions of $\alpha_s$. Measurements of both the integrated jet shape and mean 
subjet multiplicity in inclusive jet NC DIS were made. The lowest non-trivial-order 
contribution to these quantities is given by $\mathcal{O} (\alpha \alpha_s)$ pQCD 
calculations. Thus measurements of the jet shape and subjet multiplicity provide a stringent 
test of pQCD calculations beyond LO and allow a determination of $\alpha_s$. Subjets 
were resolved within a jet by considering all particles that are associated with it 
and by repeating the application of the $k_T$ cluster algorithm~\cite{kt} until, for every pair 
of particles $i$ and $j$, the quantity $d_{ij}$ was above 
$d_{\rm cut} = y_{\rm cut} \cdot (E_T^{\rm jet})^2$. All remaining clusters were called 
subjets. The subjet structure depends upon the value chosen for the resolution parameter, 
$y_{\rm cut}$.

\begin{figure}[hbp]
\begin{center}
~\epsfig{file=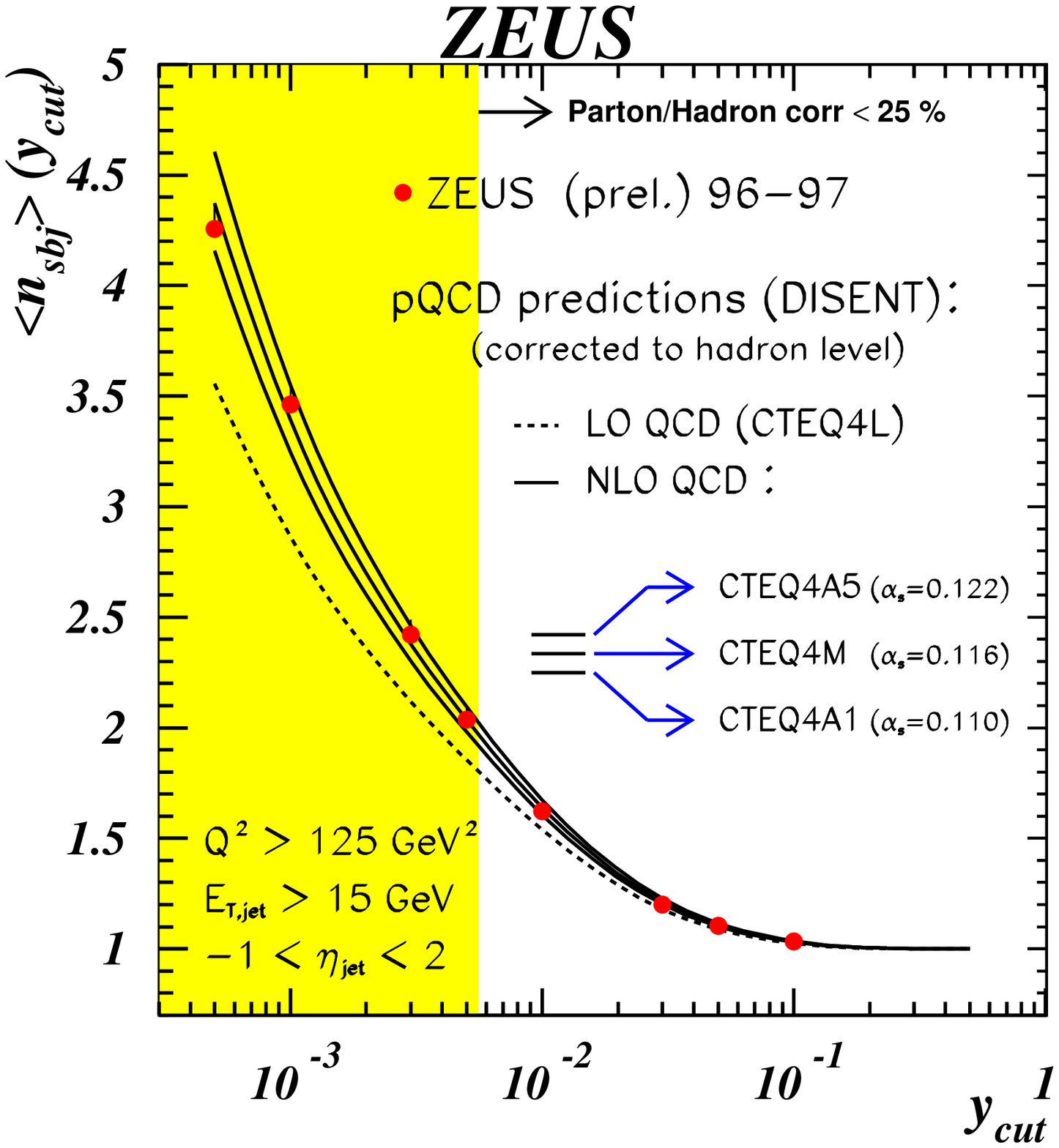,height=7.65cm}
~\epsfig{file=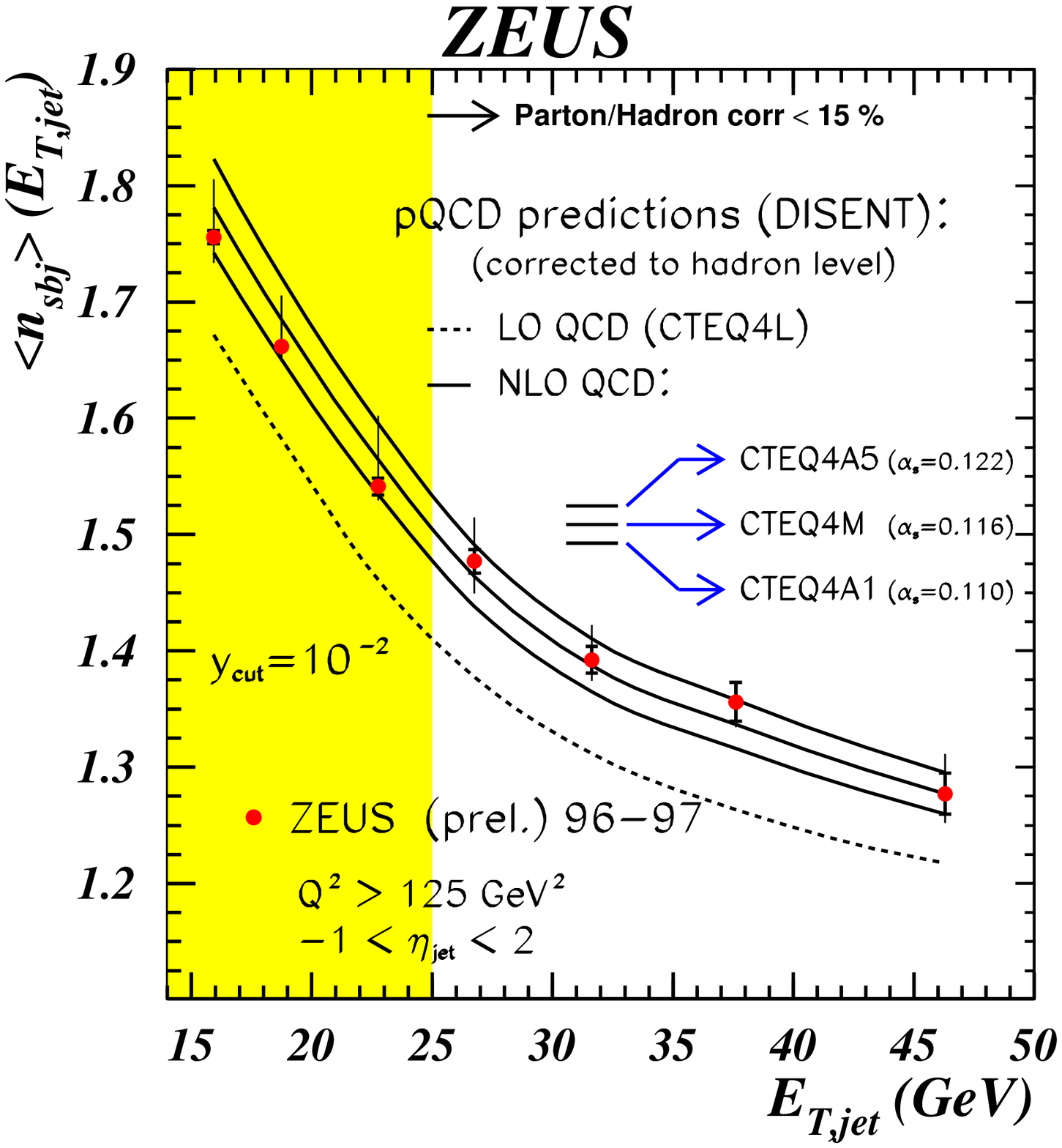,height=7.65cm}
\end{center}
\caption{\label{fig:sub_data} The mean subjet multiplicity, $<n_{\rm sbj}>$ as a function 
of $y_{\rm cut}$ and $E_T^{\rm jet}$ for fixed $y_{\rm cut}$. The data points are compared 
to pQCD predictions to LO and NLO, both corrected for hadronisation effects.}
\end{figure}

In figure~\ref{fig:sub_data}, the measurements of mean subjet multiplicity are shown 
as a function of the resolution scale, $y_{\rm cut}$, and $E_T^{\rm jet}$ at a fixed value 
of $y_{\rm cut}$ equal to 0.01. Also shown are the predictions from LO and NLO QCD, corrected 
for effects of hadronisation. The NLO predictions give a good description of the data 
for all values of $y_{\rm cut}$ and $E_T^{\rm jet}$ and can therefore be used to perform 
a measurement of $\alpha_s$. Similar results are also seen for the measurements of the 
jet shape.


\section{The gluon density in the proton}

As stated in section~\ref{sec:intro}, jet production at HERA is directly sensitive to the 
gluon content of the proton at LO. Therefore, the measurements shown here can in principle 
be used to constrain the gluon density, complementing the extractions from (predominantly) 
measurements of the proton structure function, $F_2^p$. This has been performed by 
simultaneously fitting the inclusive and dijet cross sections and the measurements from 
inclusive DIS. The result of this fit are shown in figure~\ref{fig:h1_gluon}, where the 
extraction is compared to other parametrisations of the gluon PDF. The extraction from 
this simultaneous fit is consistent with all other parametrisations shown. The effect of 
the jet data in further constraining the gluon density would be enhanced by using the 
data at lower $Q^2$. However, due to the large theoretical uncertainties, arising 
predominantly from varying the value of the renormalisation scale, an accurate extraction 
of the gluon density from jet data remains limited. Also shown in figure~\ref{fig:h1_gluon} 
is the result of simultaneously constraining both the gluon density and measuring $\alpha_s$. 
The data are sensitive to the product $\alpha_s \cdot xg(x)$, but do not permit a precise 
determination of both simultaneously.

\begin{center}
~\epsfig{file=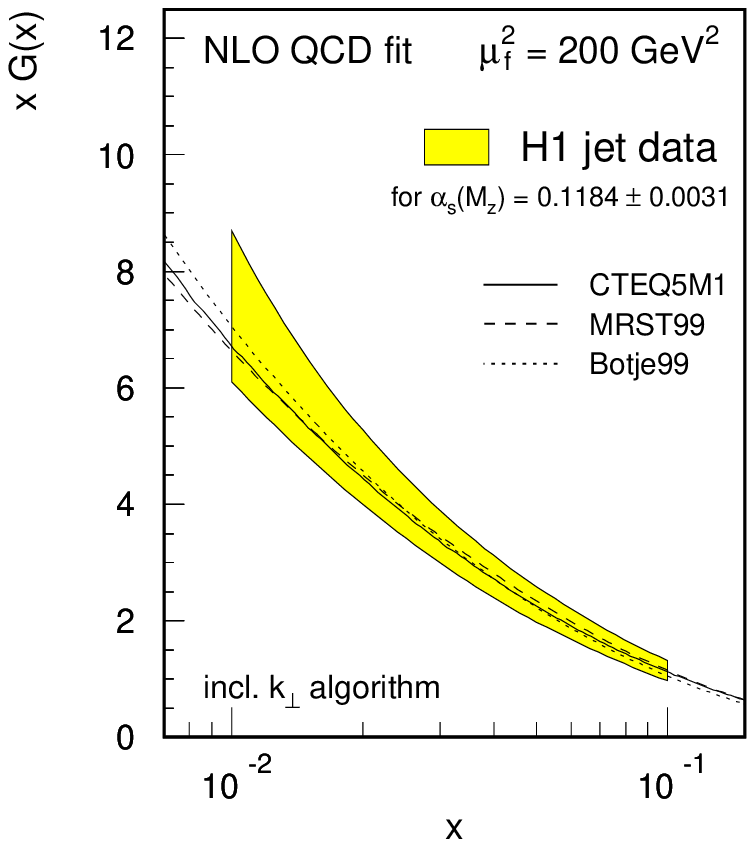,height=8cm}
~\epsfig{file=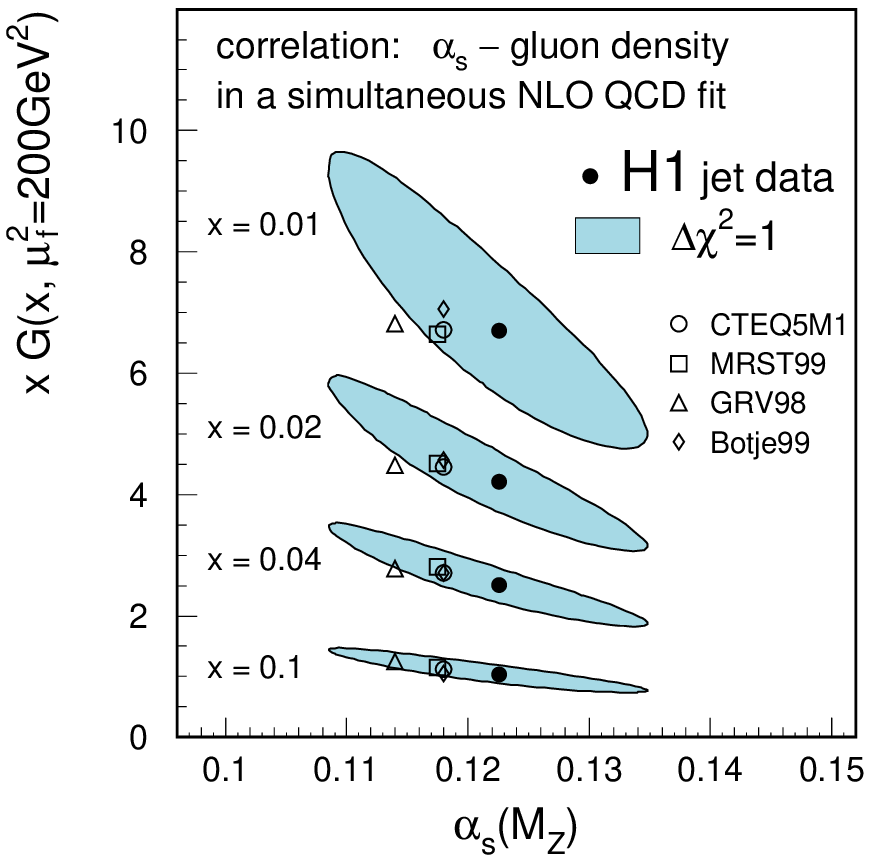,height=8cm}
\end{center}
\unitlength=1mm
\begin{picture}(0,0)(100,100)
\put(157,140){\bf (a)}
\put(242,140){\bf (b)}
\end{picture}
\begin{figure}[hbp]
\vspace{-1.cm}
\caption{\label{fig:h1_gluon} (a) The gluon density, $xG(x)$, in the proton, determined in a 
simultaneous fit to the inclusive-jet and dijet cross sections and the inclusive DIS cross 
section. The error band includes experimental and theoretical uncertainties. (b) Correlation 
of simultaneously constraining the gluon density and measuring $\alpha_s$.}
\end{figure}

\section{Measurements of $\alpha_s$}

As shown in the previous sections, measurements of jet cross sections and jet substructure 
provide an opportunity 
to extract a value for $\alpha_s$. There now exist several of these measurements from HERA as 
well as measurements of $\alpha_s$ from inclusive DIS data. At HERA there is also the 
opportunity to make a measurement of $\alpha_s$ at different values of the scale, $Q$ 
or $E_T$ and hence test the running of the value. This is shown in 
figure~\ref{fig:hera_alphas}(a), where the running of $\alpha_s$ is seen over a wide 
range in $E_T$ (within one experiment), consistent with the renormalisation group equation. 
The values of the different methods of 
measuring $\alpha_s$ are shown in figure~\ref{fig:hera_alphas}(b) and compared with the 
world averages. All the values measured are consistent both with each other and the world 
averages. In particular, the uncertainties on the measurements are also competitive with those 
on the world average. It will be interesting to see the impact these HERA measurements 
have on future determinations of the world average.

\begin{center}
~\epsfig{file=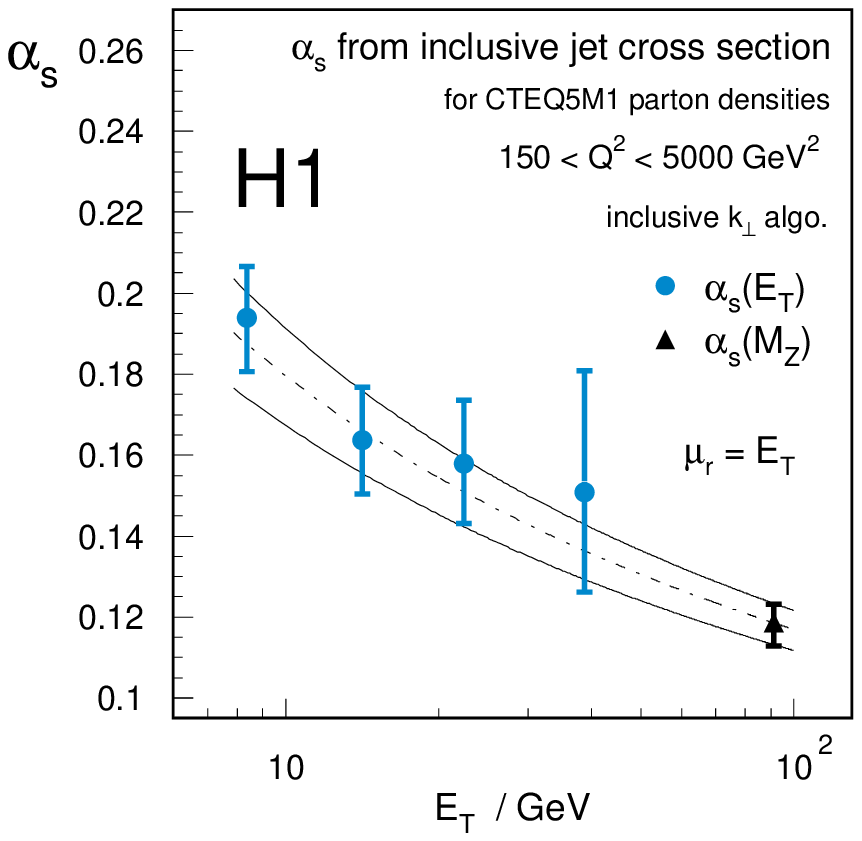,height=8.3cm}
~\epsfig{file=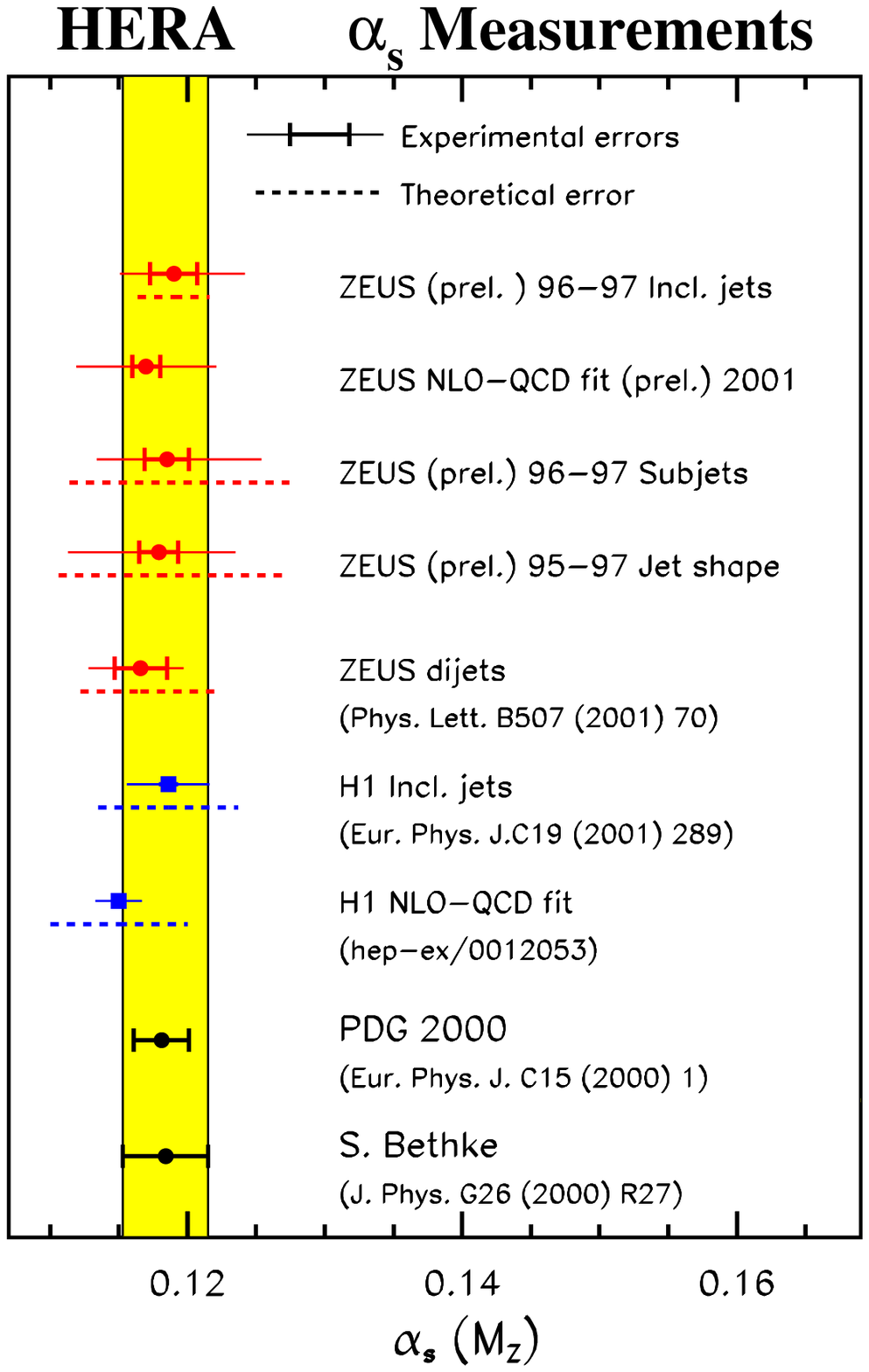,height=8.7cm}
\end{center}
\unitlength=1mm
\begin{picture}(0,0)(100,100)
\put(130,128){\bf (a)}
\put(242,120){\bf (b)}
\end{picture}
\begin{figure}[hbp]
\caption{\label{fig:hera_alphas} (a) Measurement of $\alpha_s$ as a function of the jet 
transverse energy. (b) Compilation of $\alpha_s$-measurements from HERA compared to the 
world averages.}
\end{figure}

The measurements of the inclusive jet cross sections at high-$Q^2$ show significant 
potential for the future. The uncertainty associated with that of the renormalisation scale 
is significantly smaller than in the NLO-QCD fits to $F_2^p$ data. Using the additional 
statistics available and those of the future, these measurements could provide very 
accurate measurements of $\alpha_s$. This could also be true for the measurement of the 
three- to two-jet cross section ratio, with indications of small theoretical uncertainties 
already observed. The theoretical uncertainties will also be significantly reduced with 
higher order calculations which are expected soon for fits to $F_2^p$ data~\cite{vogt} and 
within a few years for jet calculations~\cite{glover}.

\section{Discussion and summary}

Many measurements of jet production in DIS have been made over a wide kinematic range at 
HERA. In general, the data are well described by NLO QCD calculations and the current proton 
parton densities (convoluted with a correction for hadronisation effects), particularly 
at high scales. From these data, several measurements of $\alpha_s$ have been made, all of 
which agree with the world average and have uncertainties which are also competitive with 
the world average. 

Although providing good tests of pQCD, further experimental and theoretical work is needed 
to make really precise comparisons. Most of the data discussed here, use luminosities of 
the order of 30~${\rm pb^{-1}}$. Both experiments currently have about 100~${\rm pb^{-1}}$ 
of data on tape and expect to collect about ten times this amount by the end of HERA II. This 
significant increase in the data sample will enable high-precision comparisons up to very 
high-$Q^2$ values which are currently statistically limited. Experimentally, the main source 
of systematic uncertainty arises from uncertainty in the knowledge of the jet energy scale. 
The current 
quoted uncertainty is generally $\sim 3-5\%$, which can produce an uncertainty of $\sim 20\%$ 
in the cross section. Efforts are being made (and succeeding) to reduce the uncertainty to 
$\sim 1-2\%$, which will lead to an uncertainty in the cross section of $\sim 5\%$. As 
most measurements already suffer from dominating theoretical errors, particularly for values 
of $Q^2$ less than 500~${\rm GeV^2}$, these further experimental improvements will require 
much more accurate theoretical predictions. At low-$Q^2$, the renormalisation scale uncertainty 
is often greater than $50\%$, which is too large and hinders conclusions being drawn from the 
comparison with data, such as the existence of BFKL effects and the need for a resolved 
photon. Also in this region, the choice of scale is unclear with that of $E_T$ seeming to 
be the natural choice, as $E_T >> Q$, but with $Q$ appearing to better describe the data.
At high-$Q^2$, the renormalisation scale uncertainty is still the dominant source of 
error and its reduction essential for future measurements. There is a strong need for higher 
order calculations or resummed NLO programs to fully exploit the potential of the measurements 
being made.

\end{document}